\documentstyle[aps,amsfonts,amssymb]{revtex}
\begin{document}
\title{Semigroup Representations
of the Poincar\'e Group and Relativistic Gamow Vectors.}
\author{A.~Bohm, H.~Kaldass and S.~Wickramasekara}
\address{Department of Physics, University of Texas at Austin}
\author{P.~Kielanowski}
\address{Departamento de F\'{\i}sica, Centro de Investigaci\'on
y de Estudios Avanzados del IPN, Mexico City}
\date{\today}
\maketitle
\begin{abstract}
Gamow vectors are generalized
eigenvectors (kets) of self-adjoint Hamiltonians with complex
eigenvalues $(E_{R}\mp i\Gamma/2)$ describing quasistable states.
In the relativistic domain
this leads to Poincar\'e semigroup representations which are
characterized by spin $j$ and by complex 
invariant mass square
${\mathsf{s}}={\mathsf{s}}_{R}=\left(
M_{R}-\frac{i}{2}\Gamma_{R}\right)^{2}$. 
Relativistic Gamow kets have all the properties 
required to describe relativistic resonances and quasistable particles with
resonance mass $M_{R}$ and lifetime $\hbar/\Gamma_{R}$.
\end{abstract}
\section{Introduction}
Following Wigner~\cite{Wigner}, an elementary relativistic quantum
system, an elementary particle with mass $m$ and spin $s$ is in the
mathematical theory described by the space of a unitary irreducible
representation (UIR) of the Poincar\'e group~${\mathcal P}$. From
these UIR, the relativistic quantum fields are constructed
\cite{Weinberg}.  More complicated relativistic systems are described
by direct sums of UIR (for ``towers'' of elementary particles) or by
direct products of UIR (for combination of two or more elementary
particles). A direct product of UIR may be decomposed
into a continuous direct sum (integral) of irreducible representations
\cite{Joos,Macf}.  The UIR are characterized by three invariants
$(m^2, j,\textrm{sign}(p_0))$, where~$j$ represents the spin and the
real number $m$ represents the mass of elementary particle (we
restrict ourselves here to $\textrm{sign}(p_0)=+1$).

The UIR of the Poincar\'e group ${\mathcal P}$ describe stable
elementary particles (stationary systems). 
The vast majority of elementary
particles 
are unstable and UIR
provide only a more or less approximate description of them.  
The meaning of unstable elementary particles, in particular
in the relativistic domain, has always been a subject of debate.
This has recently flared-up in connection with the line shape 
analysis of the $Z$-boson, where one has difficulties to agree
upon a definition of resonance mass $m$ and width $\Gamma$.
Going back to Wigner's definition of fundamental relativistic
particles we want to
present here a special class of (non-unitary) semi-group
representations of $\cal P$ which describe quasistable relativistic
particles.

Phenomenologically, one always takes the point of view that resonances
are autonomous quantum physical entities, and decaying particles are
no less fundamental than stable particles. Stable particles are not
qualitatively different from quasistable particles, but only
quantitatively by a zero (or very small) value of $\Gamma$. Therefore
both stable and quasistable states should be described on the same
footing. This has been accomplished in the non-relativistic case,
where a decaying state is described by a generalized eigenvector of
the (self adjoint, semi-bounded) Hamiltonian with a complex eigenvalue
$z_R=E_R-i\Gamma/2$~\cite{Bohm1} and exponential time
evolution, called Gamow vectors. The stable
state vectors with real eigenvalues $E_S$ are the special case with
$\Gamma=0$.
\section{Gamow vectors}
In the standard Hilbert space formulation of quantum mechanics, such
Gamow vectors can not exist and one has to employ a formulation
based on the Rigged Hilbert Space (RHS)~\cite{antoine}. 
Dirac's bras and kets are,
mathematically, generalized eigenvectors with real eigenvalues, and
Gamow vectors are generalizations of Dirac kets. They are described by
kets $\psi^G\equiv | z_R^-\rangle\sqrt{2\pi\Gamma}$ with complex
eigenvalue $z_R=E_R-i\Gamma/2$, where $E_R$ and $\Gamma$ are
respectively interpreted as resonance energy and width. 
Like Dirac
kets, the Gamow kets are functionals of a Rigged Hilbert Space :
\begin{equation}
\Phi_+\subset{\mathcal H}\subset\Phi^\times_+:
\,\,\,\,\,\,
\psi^G\in\Phi^\times_+,
\label{eq1}
\end{equation}
and the mathematical meaning of the eigenvalue equation
$H^\times|z_R^-\rangle=(E_R-i\Gamma/2)|z_R^-\rangle$ is:
\begin{equation}
\langle H\psi|z_R^-\rangle
\equiv
\langle\psi|H^\times|z_R^-\rangle=
z_R\langle\psi|z_R^-\rangle
\,\,\,\,\,
\textup{for all}
\,\,\,\,\,
\psi\in\Phi_+.
\label{eq2}
\end{equation}
The conjugate operator $H^\times$ of the Hamiltonian $H$ is uniquely
defined by the first equality in~(\ref{eq2}), as the extension of the
Hilbert space adjoint operator $H^\dagger$ to the space of functionals
$\Phi^\times_+$~\footnote{For (essentially) self adjoint $H$,
$H^\dagger$ is equal to (the closure of) $H$; but we shall use the
definition~(\ref{eq2}) also for unitary operators ${\mathcal U}$ where
${\mathcal U}^\times$ is the extension of ${\mathcal U}^\dagger$, but
not of ${\mathcal U}$.}; on the space ${\mathcal H}$, the operators
$H^\times$ and $H^\dagger$ are the same.

The non-relativistic Gamow vectors have the following properties:
\begin{enumerate}
\item
They have an asymmetric (i.e., $t\geq 0$ only) 
time evolution and obey then an exponential law:
\begin{equation}
\psi^G(t)=\textrm{e}_+^{-iH^\times t}
|E_R-i\Gamma/2^-\rangle=
\textrm{e}^{-iE_R t}
\textrm{e}^{-\Gamma t/2}
|E_R-i\Gamma/2^-\rangle,
\,\,\,\,
\text{only for}
\,\,\,\,
t\geq0.
\label{eq3}
\end{equation}
There is another Gamow vector
$\tilde\psi^G=|E_R+i\Gamma/2^+\rangle\in\Phi^\times_-$, and another
semigroup $\textrm{e}_-^{-iH^\times t}$ for $t\leq0$ in another RHS
$\Phi_-\subset{\mathcal H}\subset\Phi_-^\times$ (with the same
${\mathcal H}$) with the asymmetric evolution
\begin{equation}
\tilde\psi^G(t)=\textrm{e}_-^{-iH^\times t}
|E_R+i\Gamma/2^+\rangle=
\textrm{e}^{-iE_R t}
\textrm{e}^{\Gamma t/2}
|E_R+i\Gamma/2^+\rangle,
\,\,\,\,
\textup{only for}
\,\,\,\,
t\leq0.
\label{eq4}
\end{equation}
\item
The $\psi^G$ ($\tilde\psi^G$) is derived as a functional at the
resonance pole term located at $z_R=(E_R-i\Gamma/2)$ (at
$z^*_R=(E_R+i\Gamma/2)$) in the second sheet of the analytically
continued S-matrix.
\item
The Gamow vectors have a Breit-Wigner energy distribution
\begin{equation}
\langle^-E|\psi^G\rangle=
i\sqrt{\frac{\Gamma}{2\pi}}\frac{1}{E-(E_R-i\Gamma/2)},
\,\,\,\,
-\infty_{II}<E<\infty,
\label{eq5}
\end{equation}
where $-\infty_{II}$ means that it extends to $-\infty$ on the second
sheet of the S-matrix (whereas the standard Breit-Wigner extends to
the threshold $E=0$).
\end{enumerate}
We want to present here a generalization of these non-relativistic
Gamow vectors to the relativistic case. 

In the non-relativistic case the inclusion of the degeneracy quantum
numbers of energy, i.e., the extension of the Dirac-Lippmann-Schwinger
kets
\begin{eqnarray}
&&|E^\pm\rangle=|E\rangle+\frac{1}{E-H\pm i0}V|E\rangle
=\Omega^\pm|E\rangle\nonumber\\[6pt]
&&H|E^\pm\rangle=
E|E^\pm\rangle;
\,\,\,\,
(H-V)|E\rangle=E|E\rangle
\label{eq6}
\end{eqnarray}
to the basis of the whole Galilei group is trivial.

For the two particle scattering states (direct product of two
irreducible representations of the Galilei group~\cite{ref5}) one uses
eigenvectors of angular momentum $(jj_3)$ for the relative motion and
total momentum ${\bbox p}$ for the center of mass motion. Thus
\begin{equation}
|E^{\textrm{\scriptsize tot}}{\bbox p}jj_3(l,s)
\,\,^\pm\rangle=
|{\bbox p}\rangle\otimes|Ejj_3\,\,^\pm\rangle
\label{eq7}
\end{equation}
where $E^{\textrm{\scriptsize tot}}=\frac{{\bbox p}^2}{2m}+E$ (the
Hamiltonian in~(\ref{eq6}) is $H=H^{\textrm{\scriptsize
tot}}-\frac{{\bbox P}^2}{2m}$).

The center-of-mass motion is usually separated by transforming to
the center-of-mass frame and then ignoring the center-of-mass 
motion
$$
|{\bbox p}=\bbox{0}\rangle\otimes|Ejj_3\,\,^\pm\rangle
\rightarrow |E,jj_3\,\,^\pm\rangle\,.
$$
For the vector in~(\ref{eq6}) one then uses 
the generalized eigenvectors of $H$ and of angular momentum
\begin{equation}
|E^{\pm}\rangle=
|Ejj_3\,\,^\pm\rangle\in\Phi^\times_\mp\supset{\mathcal H}\supset\Phi_\mp
\label{eq9}
\end{equation}
with
\begin{equation}
H^\times|Ejj_3\,\,^\pm\rangle=
E|Ejj_3\,\,^\pm\rangle,\quad 0\leq E <\infty\,.
\label{eq10}
\end{equation}
The vectors~(\ref{eq9}) are the 
Dirac-Lippmann-Schwinger scattering states and 
$E$ runs along the cut on the positive real axis of the 1-st
sheet of the $j$-th partial S-matrix. The
proper eigenvectors of $H$ with $E=-|E_n|$ at the poles on the
negative real axis of the 1-st sheet are the bound states
$|E_njj_3\rangle$. By the Galilei transformation one can
transform these vectors~(\ref{eq9}) to arbitrary momentum ${\bbox p}$; $E$ and
${\bbox p}$ are not intermingled by Galilei transformations.

To obtain the non-relativistic Gamow kets one analytically continues
the Dirac-Lippmann-Schwinger ket~(\ref{eq9}) 
into the second sheet of
the $j$-th partial S-matrix to the position of the resonance pole
$|z_R=E_R-i\Gamma/2,j,j_3\ ^-\rangle$ and obtains the following
representation~\cite{Bohm1}:
\begin{equation}
\label{11}
|z_R=E_R-i\Gamma/2,j,j_3\ ^-\rangle=
\frac{i}{2\pi}\int_{-\infty_{II}}^{+\infty}dE
|E,j,j_3\ ^-\rangle\frac{1}{E-z_R}.
\end{equation}
A Galilei transformation can boost this Gamow ket to any \textit{real}
momentum~$\bbox{p}$
\[
|{\bbox p},z_R,jj_3\ ^-\rangle=
{\mathcal U}(\bbox{p})|\bbox{0}\rangle\otimes|z_Rjj_3\ ^-\rangle.
\]
Complex momenta cannot be obtained in this way since the
Galilei transformations commute with the intrinsic energy
operator~$H$.
\section{Poincar\'e group representations with four velocity
basis}
In the relativistic case the Lorentz transformation -- in particular
Lorentz boosts -- intermingle energy $E^{\textrm{\scriptsize
tot}}=p^0$ and momenta $p^i$, $i=1,2,3$.  Thus if energy
and/or mass were complex, this would also lead
to complex momentum. 
To restrict the unwieldy set of
Poincar\'e group representations with complex momenta 
we will consider a special class of
``minimally complex'' irreducible representations of ${\mathcal P}$ to
describe relativistic resonances and decaying elementary
particles. Our construction will also lead to complex momenta $p^\mu$,
but in our case the momenta will be ``minimally complex'' in such a
way that the 4-velocities $\hat{p}_\mu\equiv\frac{p_\mu}{m}$ remain
real. This construction was motivated by a remark of
D.~Zwanziger~\cite{zwanzi} and is based on the fact that the
4-velocity eigenvectors $|\hat{\bbox{p}}j_{3}(mj)\rangle$ furnish as
valid a basis for the representation space of ${\mathcal P}$ as the
usual Wigner basis of momentum eigenvectors
$|\bbox{p}j_{3}(m,j)\rangle$.  This means every state 
$\phi\in\Phi\subset{\cal H}(m,j)\subset\Phi^{\times}$
of an UIR $(m^2,j)$, (where
$\Phi$ denotes the space of well-behaved vectors and $\Phi^{\times}$
the space of kets for the Hilbert space ${\cal H}(m,j)$ of an UIR),
can be written according to Dirac's basis vector decomposition as
\begin{equation}
\label{2.15a}
\phi=\sum_{j_{3}}\int \frac{d^{3}\hat{p}}{2\hat{p}^{0}}
|\hat{\bbox{p}},j_{3}\rangle\langle j_{3},\hat{\bbox{p}}|\phi\rangle
\end{equation}
where we have chosen the invariant measure
\begin{equation}
\label{measurehat}
d\mu(\hat{\bbox p}) = \frac{d^{3}\hat{p}}{2\hat{p}^{0}}
             = {\frac{1}{m^{2}}} \, {\frac{d^{3}p}{2 E({\bbox p})}},
\,\,\,\,\,\,     \hat{p}^{0} = \sqrt{1+\hat{\bbox p}^{2}} \, .
\end{equation}
As a consequence of (\ref{measurehat}), 
the $\delta$-function normalization of these velocity-basis vectors is
\begin{equation}
\langle \xi , \hat{\bbox p}\,|\,\hat{\bbox p}', \xi' \rangle
       = 2 \hat{p}^{0}  \delta^{3}(\hat{\bbox p}-\hat{\bbox p}')
                              \, \delta_{\xi \xi'}\\
 = 2 p^{0} m^{2} \delta^{3}
(\bbox{p}-\bbox{p'})\, 
\delta_{\xi \xi'} \, .
\label{normalizationhat} 
\end{equation}
Here, $|\hat{\bbox{p}},j_{3}\rangle\in \Phi^{\times}$ are the
eigenkets of the 4-velocity operator $\hat{P}_{\mu}=P_{\mu}M^{-1}$
and $|\phi_{j_{3}}(\hat{\bbox{p}})|^{2}
=|\langle j_{3}\hat{\bbox{p}}|\phi\rangle|^{2}$
represents the 4-velocity distribution of the vector $\phi$.
The 4-velocity eigenvectors are often
more useful for physical reasoning, because 4-velocities seem to
fulfill to rather good approximation ``velocity super-selection
rules'' which the momenta do not~\cite{ref7}.
Their use as basis vectors of the Poincar\'e group
representation~(\ref{2.15a}) does not constitute an approximation.

The relativistic Gamow vectors will be defined, not as
momentum eigenvectors, but as 4-velocity eigenvectors in the direct
product space of UIR spaces for the decay products of the resonance
$R$. We want to obtain the relativistic Gamow vectors from the pole
term of the relativistic S-matrix in complete analogy to the way the
non-relativistic Gamow vectors were obtained~\cite{Bohm1}. In the
absence of a vector space description of a resonance, we shall also in
the relativistic theory define the unstable particle by the pole of
the analytically continued partial S-matrix with angular momentum
$j$ at the value
${\mathsf{s}}={\mathsf{s}}_R\equiv(M_R-i\Gamma_{R}/2)^2$ of the invariant
mass square variable (Mandelstam variable)
${\mathsf{s}}=(p_1+p_2+\cdots)^2=E_R^2-{\bbox p}_R^2$, where $p_1$,
$p_2$,\ldots are the momenta of the decay products of
$R$~\cite{Eden,zimmermann}. This means that the mass $M_R$ and lifetime
$\hbar/\Gamma_R$ of the complex invariant mass
$w_{R}=(M_{R}-i\Gamma_{R}/2)=\sqrt{{\mathsf{s}}_{R}}$, in addition to spin
$j$, are the intrinsic properties that define a quasistable
relativistic particle~\footnote{Conventionally and equivalently one
often writes
\[
{\mathsf{s}}_R\equiv
M_\rho^2-iM_\rho\Gamma_\rho=
M_R^2\left(1-\frac{1}{4}
\left(\frac{\Gamma_R}{M_R}\right)^2\right)-iM_R\Gamma_R
\]
and calls
$M_\rho=M_R\sqrt{1-\frac{1}{4}\left(\frac{\Gamma_R}{M_R}\right)^2}$
the resonance mass and $\Gamma_\rho=\Gamma_R\left(1-\frac{1}{4}
\left(\frac{\Gamma_R}{M_R}\right)^2\right)^{-1/2}$ its width. 
We will see below that $M_{R}$ is the mass and $\hbar/\Gamma_{R}$,
not $\hbar/\Gamma_{\rho}$, is the lifetime.}.

In order to make the analytic continuation of the partial S-matrix
with angular momentum $j$, we need the angular momentum basis vectors
\begin{eqnarray}
&|\hat{\bbox p}j_3(wj)\rangle=
\int\frac{d^3\hat{p}_1}{2\hat{E}_1}
\frac{d^3\hat{p}_2}{2\hat{E}_2}
|\hat{\bbox p}_1\hat{\bbox p}_2[m_1m_2]\rangle
\langle\hat{\bbox p}_1\hat{\bbox p}_2[m_1m_2]|\hat{\bbox p}j_3(wj)\rangle
\label{eq11}\\
&\mbox{for any $(m_1+m_2)^2\leq w^2<\infty$ \,\,\,\,$j=0,1,\ldots$}
\nonumber
\end{eqnarray}
in the direct product space of the decay products of the resonance $R$
\begin{equation}
{\mathcal H}\equiv
{\mathcal H}(m_1,0)\otimes
{\mathcal H}(m_2,0)=
\int_{(m_1+m_2)^2}^{\infty}
d{\mathsf{s}}
\sum_{j=0}^{\infty}\oplus{\mathcal H}({\mathsf{s}},j),
\label{eq12}
\end{equation}
where $w^{2}={\mathsf s}$, the Mandelstam variable defined above.
For simplicity, we have assumed here that there are two decay products,
$R\rightarrow\pi_1+\pi_2$ with spin zero, described by the irreducible
representation spaces ${\mathcal H}^{\pi_i}(m_i,s_i=0)$~\footnote{Though
our discussions apply with obvious modifications to the general case of
\[
1+2+3+\cdots\rightarrow R_i\rightarrow
1^\prime+
2^\prime+
3^\prime+\cdots,
\]
these generalizations lead to enormously more complicated equations.}
of
the Poincar\'e group ${\mathcal P}$.

The kets $|\hat{\bbox p}j_3wj\rangle$ are eigenvectors of the 4-velocity
operators
\begin{equation}
\hat{P}_\mu=(P^1_\mu+P^2_\mu)M^{-1},\,\,\,\,
M^2=(P^1_\mu+P^2_\mu)({P^1}^\mu+{P^2}^\mu)
\label{eq13}
\end{equation}
with eigenvalues
\begin{equation}
\hat{p}^\mu=\left(
\begin{array}{c}
\hat{E}=\frac{p^0}{w}=\sqrt{1+\hat{\bbox p}^2}\\
\hat{\bbox p}=\frac{\bbox p}{w}
\end{array}
\right)
\,\,\,\,\,\,\mbox{and}\,\,\,\,
w^2={\mathsf{s}}.
\end{equation}
In~(\ref{eq11}) $|\hat{\bbox p}_{1}\hat{\bbox p}_{2}[m_{1}m_{2}]\rangle
=|\hat{\bbox p}_{1}m_{1}\rangle\otimes 
|\hat{\bbox p}_{2}m_{2}\rangle$ is the direct product basis
of ${\cal H}$ which are eigenvectors of
$\hat{P}^i_\mu$, the 4-velocity operators in the one
particle spaces
${\mathcal H}^{\pi_i}(m_i,s_i)$ with eigenvalues
$\hat{p}^i_\mu=\frac{p^i_\mu}{m_i}$.

To obtain the Clebsch-Gordan coefficients $\langle \hat{\bbox
p}_{1}\hat{\bbox p}_{2}[m_1,m_2]|\hat{\bbox p}j_{3}(wj)\rangle$
in~(\ref{eq11}), one follows the same procedure as given in the
classic papers~\cite{Joos,Macf,Wight,michel} for the Clebsch-Gordan
coefficients for the Wigner (momentum) basis. This has been done
in~\cite{Ref12}. The result is
\begin{eqnarray}
&\langle\hat{\bbox p}_1\hat{\bbox p}_2[m_1,m_2]|\hat{\bbox p}j_3(wj)\rangle=
2\hat{E}(\hat{\bbox p})\delta^3(\bbox{p}-\bbox{r})\delta(w-\epsilon)
Y_{jj_3}({\bbox e})\mu_j(w^2,m_1^2,m_2^2)
\label{eq15}\\
&\mbox{with}\,\,\,\epsilon^2=r^2=(p_1+p_2)^2,
\,\,\,r=p_1+p_2,\nonumber
\end{eqnarray}
where $\mu_j(w^2,m_1^2,m_2^2)$ is a function that fixes the
$\delta$-function ``normalization'' of $|\hat{\bbox
p}j_3(wj)\rangle$. The unit vector ${\bbox e}$ in~(\ref{eq15})
is the three component of $L^{-1}(r)q$, where $q$ is the
unit space like vector in the 2-plane defined by
$p_{1}$, $p_{2}$ and orthogonal to $r$~\cite{michel}.
In the c.m.\ frame the direction of ${\bbox e}$ is $\hat{\bbox
p}_1^{\textrm{\scriptsize cm}} =-\frac{m_2}{m_1}\hat{\bbox
p}_2^{\textrm{\scriptsize cm}}$.

The normalization of the basis vectors~(\ref{eq11}) is chosen to be
\begin{eqnarray}
&\langle\hat{\bbox p}^\prime j^\prime_3(w^\prime j^\prime)
|\hat{\bbox p}j_3(wj)\rangle=
2\hat{E}(\hat{\bbox p})\delta(\hat{\bbox p}^\prime-\hat{\bbox p})
\delta_{j_3^\prime j_3}\delta_{j^\prime j}
\delta({\mathsf{s}}-{\mathsf{s}}^\prime)
\label{eq16}\\
&\mbox{where}\,\,\,\,
\hat{E}(\hat{\bbox p})=\sqrt{1+\hat{\bbox p}^2}=\frac{1}{w}
\sqrt{w^2+\bbox{p}^2}\equiv
\frac{1}{w}E({\bbox p},w).\nonumber
\end{eqnarray}
This determines the weight function $\mu_j(w^2,m_1^2,m_2^2)$ to be
\begin{equation}
\left| \mu_j(w^2,m_1^2,m_2^2)\right|^{2}=
\frac{2m_1^2m_2^2w^2}{\sqrt{\lambda(1,(\frac{m_1}{w})^2,(\frac{m_2}{w})^2)}}
\label{eq17}
\end{equation}
where $\lambda$ is defined by~\cite{Wight}
\begin{equation}
\lambda(a,b,c)=a^2+b^2+c^2-2(ab+bc+ac).
\label{eq18}
\end{equation}
The basis vectors~(\ref{eq11}) are the eigenvectors of the free
Hamiltonian $H_0=P^1_0+P^2_0$
\begin{equation}
H_0^{\times}|\hat{\bbox p}j_3(wj)\rangle=
E|\hat{\bbox p}j_3(wj)\rangle,\,\,\,\,\,\,\,
E=w\sqrt{1+\hat{\bbox{p}}^2}=\sqrt{{\mathsf s}(1+\hat{\bbox p}^2)}.
\label{eq19}
\end{equation}
The Dirac-Lippmann-Schwinger scattering states are obtained, in
analogy to~(\ref{eq6}) (cf. also~\cite{Weinberg} sect.~3.1) by:
\begin{equation}
|\hat{\bbox p}j_3(wj)^\pm\rangle=
\Omega^\pm|\hat{\bbox p}j_3(wj)\rangle
\label{eq20}
\end{equation}
where $\Omega^\pm$ are the M{\o}ller operators. For the basis vectors
at rest, (\ref{eq20}) is given in analogy to~(\ref{eq6})
by the solution of the
Lippmann-Schwinger equation
\begin{equation}
|\bbox{0}j_3(wj)^\pm\rangle=
\left(1+
\frac{1}{w-H\pm i\epsilon}V
\right)
|\bbox{0}j_3(wj)\rangle.
\label{eq21}
\end{equation}
They are eigenvectors of the exact Hamiltonian $H=H_0+V$
\begin{equation}
H|\bbox{0}j_3(wj)^\pm\rangle=
\sqrt{{\mathsf{s}}}|\bbox{0}j_3(wj)^\pm\rangle,
\,\,\,\,
(m_1+m_2)^2\leq {\mathsf{s}}<\infty.
\label{eq22}
\end{equation}
The basis vectors $|\hat{\bbox p}j_3
({\mathsf{s}}j)^\pm\rangle$ of the UIR $({\mathsf{s}},j)$ 
are obtained from the basis
vectors at rest $|\bbox{0}j_3(wj)^\pm\rangle$ by the boost
(rotation-free Lorentz transformation) ${\mathcal U}(L(\hat{p}))$
whose parameters are the 4-velocities $\hat{p}^{\mu}$. The generators 
of the Lorentz transformations are the
interaction-incorporating observables
\begin{equation}
P_0=H,
\,\,\,\,
P^m,
\,\,\,\,
J_{\mu\nu},
\label{eq23}
\end{equation}
i.e., the exact generators of the Poincar\'e group (\cite{Weinberg}
sec.\ 3.3). These vectors $|\hat{\bbox p}j_{3}({\mathsf{s}}j)^{\pm}\rangle$
in (\ref{eq20}), or $|\bbox{0}j_{3}(wj)^{\pm}\rangle$ in (\ref{eq21})
when boosted by ${\mathcal U}(L(\hat{p}))$ or precisely 
${\mathcal U}^{\times}(L(\hat{p}))$, 
span the unitary representation space of the Poincar\'e
group~(\ref{eq12}) with the ``exact generators''~(\ref{eq23}).
We will be use these Dirac-Lippmann-Schwinger kets to
define the relativistic Gamow kets by analytic continuation.
\section{Relativistic Gamow kets}
The relativistic Gamow kets are defined from the Dirac-Lippmann-Schwinger
kets~(\ref{eq21}) or~(\ref{eq20}) by contour integrals around the poles of the
$j$-th partial $S$-matrix element. Starting with the
$S$-matrix element
\begin{equation}
\label{26a}
(\psi^{out},S\phi^{in})=(\psi^-,\phi^+)=
\sum_{jj_3} \int_{(m_1+m_2)^{2}}^{\infty}
d{\mathsf{s}}\int d\mu(\hat{\bbox p})
\langle\psi^-|\hat{\bbox p}j_3{\mathsf{s}}j^-\rangle S_j({\mathsf{s}})
\langle^+\hat{\bbox p}j_{3}{\mathsf{s}}j|\phi^+\rangle
\end{equation}
one deforms the contour of integration over ${\mathsf{s}}$ from the
physical values $(m_1+m_2)^{2}\leq{\mathsf{s}}<\infty$
on the upper rim of the cut along the ${\mathsf{s}}$-axis, into the 
second sheet past the pole at ${\mathsf{s}}_{R}$. For the integration around
the pole ${\mathsf{s}}_{R}$ the integral~(\ref{26a}) splits
of a pole term which defines the Gamow vector 
$|\hat{\bbox p}j_3{\mathsf{s}}_{R}j^-\rangle$. This is done in exactly
the same way as in the non-relativistic case~\cite{Bohm1} and leads to the 
relativistic analogue of~(\ref{11})~:
\begin{equation}
\label{28}
\langle \psi^{-}|\hat{\bbox p}j_{3},{\mathsf s}_{R}j^{-}\rangle
\equiv-\frac{i}{2\pi}\oint d{\mathsf s}\langle\psi^{-}|
\hat{\bbox p}j_{3}{\mathsf s}j^{-}\rangle\frac{1}{{\mathsf s}-
{\mathsf s}_{R}}=
\frac{i}{2\pi}
\int_{-\infty_{II}}^{+\infty_{II}}d{\mathsf{s}}
\langle\psi^{-}|\hat{\bbox p}j_3{\mathsf{s}} j^{-}
\rangle\frac{1}{{\mathsf{s}}-{\mathsf{s}}_R}\quad
\text{ for all }\psi^{-}\in\Phi_{+}\,.
\end{equation}
For this analytic continuation to be possible the RHS formulation
of quantum theory makes a new hypothesis~:\\
The set of prepared in-states $\{\phi^+\}$ and the set of detected
out-states (decay products) $\{\psi^-\}$ form two different dense
subspaces of the Hilbert space ${\cal H}$, cf.~(\ref{eq12}) 
and therewith two distinct RHS's
\begin{mathletters}
\label{rhs}
\begin{eqnarray}
\label{rhs1}
&\Phi_-\subset{\cal{H}}\subset\Phi_{-}^{\times}\quad
\text{for prepared in-states }\phi^+\\
\label{rhs2}
&\Phi_+\subset{\cal{H}}\subset\Phi_+^{\times}\quad
\text{for detected out-states (observables) }\psi^-\,.
\end{eqnarray}
\end{mathletters}
where $\Phi_-\,(\Phi_+)$ is of the Hardy class type in the lower (upper)
half plane. This means $\langle^-\hat{\bbox p}j_3{\mathsf{s}}j|\psi^-\rangle$
($\langle^+\hat{\bbox p}j_3{\mathsf{s}}j|\phi^+\rangle$)
are well behaved Hardy class functions of the variable ${\mathsf{s}}$
in the upper (lower) half plane second sheet. This new hypothesis,
which distinguishes meticulously between states (accelerator) and observables
(detector) was justified in the non-relativistic case by some causality
arguments~\cite{antoniou}. All our new results can be derived from this 
new Hardy class hypothesis which is different from the conventional
assumptions of scattering theory $\{\phi^+\}=\{\psi^-\}(={\cal H})$.

The first equality in~(\ref{28}) is the definition that
associates $\psi^{G}$ to the pole term in the second sheet, and
the second equality is a consequence of the 
Hardy class property~\cite{hardy}. As a consequence the wave function
$\langle\hat{\bbox p}j_{3}{\mathsf s}j^{-}|\psi^{G}\rangle$
of the Gamow ket is a Breit-Wigner function of ${\mathsf s}$ that
extends over all physical values of ${\mathsf s}$ and 
the non-physical values of ${\mathsf s}$ on the second sheet
to $-\infty$.

The Lorentz transformations $\Lambda$ are represented by unitary
operators ${\mathcal U}^\dagger(\Lambda)$ in ${\cal{H}}$. This means its
conjugate operator $\left({\cal U}^{\dagger}(\Lambda)\right)^\times$ 
(usually denoted as ``${\cal U}(\Lambda)$'')
acts in the space
$\Phi^{\times}_+$ in the standard way~:
\begin{eqnarray}
\label{eq25}
&\langle{\cal U}^\dagger(\Lambda)\psi^-|\hat{\bbox p}j_3,
{\mathsf s}_Rj^-\rangle=\langle\psi^-|
\left({\mathcal U}^{\dagger}(\Lambda)\right)^\times
|\hat{\bbox p}j_3,{\mathsf{s}}_R j^{-}\rangle=
\sum_{j_3^\prime}
\langle\psi^-|\bbox{\Lambda}\hat{\bbox{p}}j_3^\prime,{\mathsf{s}}_R j^{-}\rangle
D^{j}_{j_{3}^{\prime}j_{3}}
({\mathcal R}(\Lambda,\hat{p}))\\
&\,\,\text{for all }\psi^-\in\Phi_+\subset{\cal H}\,,\text{ all }
\Lambda\in SO(3,1)\,,
\nonumber
\end{eqnarray}
where ${\mathcal R}(\Lambda,\hat{p})=L^{-1}(\Lambda \hat{p})\Lambda
L(\hat{p})$ is the Wigner rotation.  
The ${\cal U}^\dagger(\Lambda)$ in~(\ref{eq25}) is the 
restriction of the unitary ${\cal U}^\dagger(\Lambda)$ to 
the dense subspace $\Phi_+$, which
remains invariant under the action of ${\cal U}(\Lambda)$
for all $\Lambda\in SO(3,1)$.
For the rotation free Lorentz boost one obtains in particular
\begin{equation}
\left({\mathcal U}^{\dagger}(L(\hat{p}))\right)^{\times}
|\hat{\bbox p}={\bbox 0},j_3,{\mathsf{s}}_R j^{-}\rangle=
|\hat{\bbox p}j_3,{\mathsf{s}}_R j^{-}\rangle\,,
\label{eq26}
\end{equation}
where the boost $L^{\mu}_{\,\,\,\nu}$ is a function 
of the real parameters $\hat{p}^\mu$ and not
of the complex $p^\mu$:
\begin{equation}
L^\mu_{\hphantom{\mu}\nu}=
\left(
\begin{array}{cc}
\frac{p^0}{m}&-\frac{p_n}{m}\\
\frac{p^k}{m}&\delta^k_n-\frac{\frac{p^k}{m}\frac{p_n}{m}}{1+\frac{p^0}{m}}
\end{array}
\right),
\,\,\,\,
L(\hat{p})
\left(
\begin{array}{c}
1\\
0\\
0\\
0
\end{array}
\right)
=\hat{p}.
\label{eq27}
\end{equation}
Thus in these representations the velocities $\hat{p}^{\mu}$
are real and the momenta $p^{\mu}=m\hat{p}^{\mu}$
become complex only through the complex factor $m=\sqrt{{\mathsf s}_{R}}$.
It is this property that leads to the semigroup representations.

The relativistic Gamow kets~(\ref{28}) are generalized eigenvectors
of the invariant mass squared operator $M^2=P_\mu P^\mu$ with
eigenvalue ${\mathsf{s}}_R$ as can be seen immediately
by using in~(\ref{28}) $M^{2}\psi^{-}\in\Phi_{+}$
in place of $\psi^{-}$
\begin{equation}
\langle\psi^-|M^{2^{\times}}
|\hat{\bbox p}j_3,{\mathsf{s}}_Rj^{-}\rangle=
\frac{i}{2\pi}\int_{-\infty}^{+\infty}d{\mathsf s}\,{\mathsf s}
\langle\psi^{-}|\hat{\bbox p}j_{3}{\mathsf s}j^{-}\rangle
\frac{1}{{\mathsf s}-{\mathsf s}_{R}}=
{\mathsf{s}}_R\langle\psi^-|\hat{\bbox p}j_3,
{\mathsf{s}}_R j^{-}\rangle
\quad \text{for every } \psi^-\in\Phi_+\subset
{\mathcal H}\subset\Phi^\times_+.
\label{32}
\end{equation}
To prove~(\ref{32}) one needs to
use the properties of the Hardy class space~\cite{hardy}.
Similarly one shows that the 
$|\hat{\bbox p}j_{3},{\mathsf s}_{R}j^{-}\rangle$ are generalized
eigenvectors of the momentum operators of~(\ref{eq23})~\cite{hardy}
\begin{equation} \label{32a}
\langle P_{\mu}\psi^-|\hat{\bbox p}j_3,{\mathsf s}_{R}j^-\rangle =
\langle \psi^{-}|P_{\mu}^{\times}|\hat{\bbox p}j_{3},{\mathsf s}_{R}
j^{-}\rangle
=\frac{i}{2\pi}\int_{-\infty}^{+\infty}
d{\mathsf s}\sqrt{{\mathsf s}}\hat{p}_{\mu}
\langle\psi^{-}|\hat{\bbox p}j_{3}{\mathsf s}j^{-}\rangle
\frac{1}{{\mathsf s}-{\mathsf s}_{R}}
=\sqrt{{\mathsf s}_{R}}\hat{p}_{\mu}
\langle\psi^{-}|\hat{\bbox p}j_{3},{\mathsf s}_Rj^{-}\rangle\, .
\end{equation}
Thus the generalized momentum eigenvalues are
``minimally complex'' $p_{\mu}=\sqrt{{\mathsf s}_{R}}\hat{p}_{\mu}$.\\
The continuous linear combinations of the 4-velocity kets (\ref{eq20}) with an
arbitrary 4-velocity distribution function
$\phi_{j_{3}}(\hat{\bbox{p}})\in {\cal S}$ (Schwartz space),
\begin{equation}
\label{32b}
\psi^{\rm G}_{j {\mathsf{s}}_{R}}=
\sum_{j_{3}}\int \frac{d^{3}\hat{p}}{2\hat{p}^{0}}
|\hat{\bbox{p}}j_{3},{\mathsf{s}}_{R}j^{-}
\rangle \phi_{j}(\hat{\bbox{p}}),
\end{equation}
also represent relativistic Gamow states with the complex
mass ${\mathsf{s}}_{R}=(M_{R}-i\Gamma_{R}/2)^{2}$.

In contrast to the action of the Lorentz subgroup~(\ref{eq25}),
the translation subgroup ${\cal U}^\dagger(x,{\bbox 1})=e^{iP^\mu x_\mu}$
does not leave the subspace $\Phi_+$ of ${\cal H}$ invariant.
However there is a semigroup of time-like translations 
$(x^+,{\bbox 1})$ into the forward light cone with
$\hat{p}^{\mu}x_{\mu}=(1+\hat{\bbox p}^{2})^{1/2}x^0-\hat{\bbox p}.{\bbox x}
\geq 0$ whose (restrictions to $\Phi_+$ of) ${\cal U}^\dagger(x^+,{\bbox 1})$
leave the subspace $\Phi_+$ invariant. The 
$\{(x^+,\Lambda)\}|\Lambda\in SO(3,1),\text{ }x_\mu \text{with }
\left( (1+\hat{\bbox p}^2)^{1/2}x^0-\hat{\bbox p}.\hat{\bbox x}\right)
\geq 0\,,\,\hat{\bbox p}\in{\mathbb R}^3\}={\cal P}_+$ form a 
semigroup and their representatives ${\cal U}^\dagger(x^+,{\bbox 1})$ are 
continuous operators on $\Phi_+$,
${\cal U}^\dagger(x^+,\Lambda)\Phi_+\rightarrow\Phi_+$.
For the other $(x,\Lambda)\in{\cal P}$ this is not fulfilled,
cf.\ the analogy to the non-relativistic case~\cite{Bohm1}.

For the particular case $\hat{\bbox p}={\bbox 0}$, $x^0=t\geq 0$
we obtain the time translation into the forward direction generated
by the energy operator $H=P_0$
\begin{eqnarray}
\label{33}
&\langle\psi^-|\textrm{e}^{-iH^{\times}t}
|\hat{\bbox p}=\bbox{0},j_3,{\mathsf{s}}_R j^{-}\rangle=
\textrm{e}^{-i\sqrt{{\mathsf s}_R}t}\langle\psi^-|\hat{\bbox p}
={\bbox 0},j_3,{\mathsf s}_R j^-\rangle=
\textrm{e}^{-iM_R t}
\textrm{e}^{-\Gamma_{R} t/2}
\langle \psi^-|\hat{\bbox p}=\bbox{0},j_3,{\mathsf{s}}_R j^{-}\rangle
\,\,\,\,\\
&\text{for all }\psi^-\in\Phi_+\text{ and }
\text{for }t\geq0\,\,\,\textrm{only}\,,
\nonumber
\end{eqnarray}
where $t$ is time in the rest system.

Thus relativistic Gamow states are representations of ${\mathcal P}_+$
with spin $j$ and complex mass
${\mathsf{s}}_R=(M_{R}-i\Gamma_{R}/2)^2\equiv m_\rho^2-im_\rho\Gamma_\rho$, for
which the Lorentz subgroup is unitarily represented. They are obtained
from the resonance pole of the relativistic partial S-matrix
$S_{j}({\mathsf{s}})$, 
and thus lead to a representation of the $j$-th partial 
scattering amplitude
\begin{mathletters}
\label{34}
\begin{equation}
\label{34a}
a_{j}({\mathsf s})=a_{j}^{\text{BW}}({\mathsf s})+B({\mathsf s})\,,
\end{equation}
where $a_{j}^{\text{BW}}({\mathsf s})$ is a relativistic
Breit-Wigner amplitude given by
\begin{equation}
a_j^{\text{BW}}({\mathsf s})
=\frac{\Gamma}{{\mathsf{s}}-(M_{R}-i\frac{\Gamma_{R}}{2})^2},
\,\,\,\,
-\infty_{II}<{\mathsf{s}}<+\infty_{II}\,,
\label{34b}
\end{equation}
\end{mathletters}
and $B({\mathsf s})$ is a background term not associated
to the resonance pole at ${\mathsf s}_{R}$. The background
is slowly varying in the neighborhood of the resonance
peak $\left(M_{R}^{2}-\Gamma_{R}^{2}/4\right)$ of
$|a_{j}^{\text{BW}}({\mathsf s})|^{2}$, unless there
is another resonance in the same partial wave at
a nearby ${\mathsf s}_{R_{2}}$ in which case the
resonance at ${\mathsf s}_{R_{2}}$ has to be treated in the 
same way and leads to 
$B({\mathsf s})\rightarrow a_{j}^{\rm{BW}_{2}}
({\mathsf s})+B'({\mathsf s})$.
\section{Summary}
The Gamow vector obeys an exact exponential decay law with a 
lifetime $\tau_{R}$ given precisely by $\tau_{R}=\hbar/\Gamma_{R}$,
according to~(\ref{33}), and not by $\hbar/\Gamma_{\rho}$
or any other $\Gamma$.
The separation~(\ref{34a}) of an exact Breit-Wigner~(\ref{34b})
and the isolation of an exactly exponential decaying Gamow
state $\psi^{G}$ associated to each Breit-Wigner of
each $S$-matrix pole
is achieved by the hypothesis~(\ref{rhs}) of the Hardy class spaces.
Only for the Gamow ket~(\ref{28}) can one prove~(\ref{33})
which leads to the exact
exponential decay law for the decay rate~\cite{harshmann}
and therewith to the precise relation $\tau_{R}=\hbar/\Gamma_{R}$.
Without the postulate~(\ref{rhs})
this {\em cannot} be derived, though it has always been assumed on the 
basis of some ``approximate'' derivations~\cite{goldberger}.
The Gamow vector also helps to decide the debate about the
right definition of the $Z$-boson mass and width~\cite{riemann}.
According to~(\ref{33}) it is probably $M_R$ and certainly $\Gamma_R$
(if one wants $\tau_R=\hbar/\Gamma_R$ to hold) which should
be called the mass and width, not the peak position $M_\rho$
of the Breit-Wigner~(\ref{34b}) and not
$M_Z=\sqrt{M_R^2\left(1+\frac{3}{4}\left(\frac{\Gamma_Z}{M_Z}\right)\right)}
+0\left(\left(\frac{\Gamma_Z}{M_Z}\right)^4\right)$
of the on-mass-shell definition.

The above are all features which one may welcome or 
easily accept for states that
are to describe relativistic resonances. In addition, 
Gamow vectors have a
semigroup time evolution $t\geq 0$~(\ref{33}),
expressing irreversibility on the microphysical level.
This may be puzzling and disturbing to many, 
but a fundamental time asymmetry of quantum physics has been
noticed independently and in more general contexts~\cite{gellmann,haag}.
The Gamow kets can represent the ``causal links'' between two
events~\cite{haag} and for microphysical ``states'' representing
causal links a semigroup time evolution is quite natural.
\section*{Acknowledgement}
We gratefully acknowledge valuable support from the 
Welch Foundation.
P.~Kielanowski gratefully acknowledges support from CoNaCyT (Mexico).


\begin{thebibliography}{99}
\bibitem{ref5}
The projectively equivalent irreducible representations $(m,E,s)\sim
(m,0,s)\equiv [m,s]$ of the Galilei group are characterized by
non-relativistic mass $m$ and spin $s$. In the direct product of
$[m_1,s_1]\otimes[m_2,s_2]$, the two spins are combined (using $SO(3)$
Clebsch-Gordan coefficients) to give the total spin $s$. Then, using
orbital angular momentum $l$ for the relative orbital motion, one
couples $s$ with $l$ to form the total angular momentum $j$ with the
3-component $j_3$. This gives the basis vectors
$|E^{\textrm{\scriptsize tot}},{\bbox p}jj_3(s,l)\rangle$ 
defined by their scalar
products in the center of mass frame ${\bbox p}={\bbox 0}$ (Clebsch-Gordan
coefficients of the Galilei group) with the basis vector
$|{\bbox p}_1,s_1,(s_1)_3\rangle\otimes
|{\bbox p}_2,s_2,(s_2)_3\rangle=
|{\bbox p}_1,(s_1)_3,{\bbox p}_2,(s_2)_3(s_1,s_2)\rangle$
\begin{eqnarray}
\langle{\bbox p}_1,(s_1)_3,{\bbox p}_2,(s_2)_3(s_1,s_2)|
E^{\textrm{\scriptsize tot}},{\bbox p}jj_3(s,l)\rangle
&=&N\delta^3({\bbox{p}-\bbox{p}_1-\bbox{p}_2})
\delta(E^{\textrm{\tiny tot}}-\frac{p_1^2}{2m_1}
-\frac{p_1^2}{2m_1})\nonumber\\
&&\times\sum_{s_3,l_3}
\langle s_1(s_1)_3s_2(s_2)_3|ss_3\rangle
\langle ll_3ss_3|jj_3\rangle
Y_{ll_3}(\bbox{e})\nonumber
\end{eqnarray}
where $\bbox{e}$ is the unit vector in the direction
of $\bbox{p}$ and $N$ is a suitable normalization factor.
\bibitem{hardy}
The proof of~(\ref{32}),~(\ref{32a}) and of the second equality
in~(\ref{28})
is a consequence of the following theorem (Titchmarsh) for Hardy
class functions $G_{-}({\mathsf s})\in{\cal H}_-^2$~:\\
For $G_{-}({\mathsf s})\in{\cal H}_{-}$ (Hardy class in the lower half plane)
and $\rm{Im}\,z>0$ one has
$$
G_{-}(z)=\frac{-1}{2\pi i}\int_{-\infty}^{+\infty}
\frac{G_{-}({\mathsf s})}{{\mathsf s}-z} d{\mathsf s}\,.
$$
$\langle \phi^+|{\mathsf s}^+\rangle\,,\text{ }\langle^-{\mathsf s}|\psi^-
\rangle\in{\cal H}_+$ are well-behaved Hardy-class functions
in the upper half plane and 
$\langle^+{\mathsf s}|\phi^+\rangle\,,\text{ }
\langle\psi^-|{\mathsf s}^-\rangle\in{\cal H}_-$
are well-behaved Hardy-class functions in the lower half plane.
This means that $\sqrt{{\mathsf s}}\langle \psi^-|{\mathsf{s}}^-\rangle$,
${\mathsf s}\langle\psi^-|{\mathsf s}^-\rangle$, $\cdots$,
${\mathsf s}^{\frac{n}{2}}\langle\psi^-|{\mathsf s}^-\rangle$, $\cdots$, 
$n=1,2,3,\cdots$ are also elements of ${\cal H}_{-}$. 
Choosing $G_{-}({\mathsf s})={\mathsf s}^{\frac{n}{2}}\langle
\psi^-|{\mathsf s}^-\rangle$ one obtains~(\ref{28}),~(\ref{32}) and~(\ref{32a}) 
etc.
\end{thebibliography}
\end{document}